\documentclass[11pt]{article}

\usepackage[reqno]{amsmath}
\usepackage{epsfig}
\usepackage{array}
\usepackage{float}

\usepackage{a4}
\usepackage{epsfig}
\usepackage{a4wide}
\newcommand{\beq}{\begin{equation}}
\newcommand{\eeq}{\end{equation}} 
\def\gs{\mathrel{ \rlap{\raise
0.511ex \hbox{$>$}}{\lower 0.511ex \hbox{$\sim$}}}} \def\ls{\mathrel{
\rlap{\raise 0.511ex \hbox{$<$}}{\lower 0.511ex \hbox{$\sim$}}}}

\newcommand{\ba}{\begin{array}{c}}
\newcommand{\baz}{\begin{array}{cc}}
\newcommand{\bad}{\begin{array}{ccc}}
\newcommand{\bea}{\begin{equation} \begin{array}{c}}
\newcommand{\eea}{ \end{array} \end{equation}}
\newcommand{\ea}{\end{array}} 
\newcommand{\dms}{\mbox{$\Delta m^2_{\odot}$ }}
\newcommand{\dma}{\mbox{$\Delta m^2_{\rm A}$ }}

\newcommand{\ppp}{\mbox{$(+++)$}} \newcommand{\pmm}{\mbox{$(+--)$}}
\newcommand{\mpm}{\mbox{$(-+-)$}} \newcommand{\mmp}{\mbox{$(--+)$}}
\def\gtap{\mathrel{ \rlap{\raise 0.511ex \hbox{$>$}}{\lower 0.511ex
   \hbox{$\sim$}}}} \def\ltap{\mathrel{ \rlap{\raise 0.511ex
   \hbox{$<$}}{\lower 0.511ex \hbox{$\sim$}}}}
   
   \newcommand{\deltasol}{\mbox{$ \Delta m^2_{\odot}$}}
   
   \newcommand{\betabeta}{\mbox{$(\beta \beta)_{0 \nu} $}}
   \newcommand{\meffih}{\mbox{$ \left|< \! m \! > \right|^{\mathrm{IH}}$}}
   
   \newcommand{\meff}{\mbox{$\left| < \! m \! > \right|$}}
   \newcommand{\meffexp}{\mbox{$(\left| < \! m \! > \right|_{\mbox{}_{\mathrm{exp}}})_{\mbox{}_{\mathrm{MIN}}}~$}}
   \newcommand{\hbeta}{$\mbox{}^3 {\rm H}$ $\beta$-decay }
   \newcommand{\eV}{\mbox{$ \ \mathrm{eV}$}}
   \newcommand{\deltatre}{\mbox{$ \Delta m^2_{32} \ $}}
   \newcommand{\deltadue}{\mbox{$ \Delta m^2_{21} \ $}}

\newcommand{\ts}{\mbox{$\tan^2
\theta_\odot$}}
\begin{document}

\hfill{Ref. SISSA 89/2002/EP} \rightline{UCLA/02/TEP/38}
\rightline{December 2002}
\rightline{hep-ph/0212113}

\begin{center}
{\bf On the Neutrino Mass Spectrum and Neutrinoless Double-Beta Decay}

\vspace{0.4cm}

S. Pascoli~$^{a)}$, \hskip 0.2cm S. T. Petcov~$^{b,c)}$
\footnote{Also at: Institute of Nuclear Research and Nuclear Energy,
Bulgarian Academy of Sciences, 1784 Sofia, Bulgaria} ~and~
W. Rodejohann~$^{b)}$

\vspace{0.2cm}

{\em $^{a)}$Department of Physics, University of California, Los Angeles CA 90095-1547, USA\\ }
\vspace{0.2cm} {\em $^{b)}$Scuola Internazionale Superiore di Studi
Avanzati, I-34014 Trieste, Italy\\ }
\vspace{0.2cm} {\em $^{c)}$Istituto Nazionale di Fisica Nucleare,
Sezione di Trieste, I-34014 Trieste, Italy\\ }

\end{center}

\begin{abstract}

   Assuming 3-$\nu$ mixing, 
neutrino oscillation explanation 
of the solar and atmospheric neutrino 
data and of the first KamLAND results, 
massive Majorana neutrinos and
neutrinoless double-beta (\betabeta-) decay generated 
only by the (V-A) charged current weak interaction 
via the exchange of the three Majorana neutrinos,
we analyze in detail the possibility of 
determining the type of the neutrino mass spectrum
by measuring of the effective Majorana mass
\meff{} in \betabeta-decay. 
The three possible 
types of neutrino mass spectrum are considered:
i) normal hierarchical (NH) $m_1 \ll m_2 \ll m_3$,
ii) inverted hierarchical (IH),
$m_1 \ll m_2 \cong m_3$, and
iii) quasi-degenerate (QD),
$m_1 \cong m_2 \cong m_3$,
$m_{1,2,3} \gtap 0.20$ eV. 
The uncertainty in 
the measured value of \meff{}
due to the imprecise knowledge of
the relevant nuclear matrix elements
is taken into account in the analysis.
We derive the ranges of values of 
$\tan^2\theta_{\odot}$,
$\theta_{\odot}$ being the 
mixing angle which controls the 
solar neutrino oscillations,
and of the nuclear matrix element 
uncertainty factor,
for which the measurement
of \meff{} would allow one
to discriminate between
the NH and IH, NH and QD 
and IH and QD spectra. 

\end{abstract}
\newpage

\vspace{-0.4cm}
\section{Introduction}

\vspace{-0.2cm}
\hskip 1.0truecm The solar neutrino experiments
Homestake, Kamiokande, SAGE, GALLEX/GNO,
Super-Kamiokande (SK) and SNO \cite{Cl98,SKsol,SNO1,SNO2}, 
the data on atmospheric neutrinos
obtained by the Super-Kamiokande (SK) 
experiment \cite{SKatm00} and 
the results from the KamLAND 
reactor antineutrino 
experiment \cite{KamLAND},
provide very strong evidences for oscillations 
of flavour neutrinos. The evidences for solar 
$\nu_e$ oscillations into active neutrinos $\nu_{\mu,\tau}$,
in particular, 
were spectacularly reinforced 
by the first data from the SNO 
experiment \cite{SNO1}
when combined with the data from the 
SK experiment \cite{SKsol}, 
by the more recent SNO data \cite{SNO2},
and by the just published 
first results of the KamLAND 
\cite{KamLAND} experiment.
Under the rather plausible 
assumption of CPT-invariance,
the KamLAND data practically 
establishes \cite{KamLAND} the
large mixing angle (LMA)
MSW solution as unique solution
of the solar neutrino problem.
This remarkable result
brings us, after more than 
30 years of research, 
initiated by the pioneer
works of B. Pontecorvo \cite{Pont4667} and the
experiment of R. Davis et al. \cite{Davis68},
very close to a complete understanding of the 
true cause of the solar neutrino problem. 

    The interpretation of the solar and
atmospheric neutrino, and of the KamLAND
data in terms of 
neutrino oscillations requires
the existence of 3-neutrino mixing
in the weak charged lepton current 
(see, e.g., \cite{BGG99,SPWIN99}):
%%%%%%%%%%%%%%%%%%%
\begin{equation}
\nu_{l \mathrm{L}}  = \sum_{j=1}^{3} U_{l j} \, \nu_{j \mathrm{L}}~.
\label{3numix}
\end{equation}
%%%%%%%%%%%%%%%%%%%
\noindent Here $\nu_{lL}$, $l  = e,\mu,\tau$,
are the three left-handed flavor 
neutrino fields,
$\nu_{j \mathrm{L}}$ is the 
left-handed field of the 
neutrino $\nu_j$ having a mass $m_j$
and $U$ is the Pontecorvo-Maki-Nakagawa-Sakata (PMNS)
neutrino mixing matrix \cite{BPont57}. 
If the neutrinos with definite mass $\nu_j$
are Majorana particles, 
the process of neutrinoless double-beta  
(\betabeta-) decay
will be allowed (for reviews see, 
e.g., \cite{BiPet87,ElliotVogel02}).
If the \betabeta-decay is generated
{\it only by the (V-A) charged current 
weak interaction via the exchange of the three
Majorana neutrinos  $\nu_j$ and the latter
have masses not exceeding few MeV},
the dependence of the 
\betabeta-decay amplitude 
on the neutrino mass and mixing parameters
factorizes in the 
effective Majorana mass \meff{}
(see, e.g., \cite{BiPet87}):
%%%%%%%%%%%%%%%%%%%%
\begin{equation}
\meff  = \left| m_1 |U_{\mathrm{e} 1}|^2 
+ m_2 |U_{\mathrm{e} 2}|^2~e^{i\alpha_{21}}
 + m_3 |U_{\mathrm{e} 3}|^2~e^{i\alpha_{31}} \right|~,
\label{effmass2}
\end{equation}
%%%%%%%%%%%%%%%%%%%%
\noindent where 
$\alpha_{21}$ and $\alpha_{31}$ 
are two Majorana CP-violating phases
\footnote{We assume that $m_j > 0$ and that
the fields of the 
Majorana neutrinos $\nu_j$ 
satisfy the Majorana condition:
$C(\bar{\nu}_{j})^{T} = \nu_{j},~j=1,2,3$,
where $C$ is the charge conjugation matrix.}
\cite{BHP80,Doi81}.
If CP-invariance holds, 
one has \cite{LW81}
$\alpha_{21} = k\pi$, $\alpha_{31} = 
k'\pi$, where $k,k'=0,1,2,...$. In this case 
%%%%%%%%%%%%%%%%%%%%
\begin{equation}
\eta_{21} \equiv e^{i\alpha_{21}} = \pm 1,~~~
\eta_{31} \equiv e^{i\alpha_{31}} = \pm 1 
\label{eta2131}
\end{equation}
%%%%%%%%%%%%%%%%%%%%
\noindent represent the relative 
CP-parities of the neutrinos 
$\nu_1$ and $\nu_2$, and 
$\nu_1$ and $\nu_3$, respectively. 
The oscillations
between flavour neutrino
are insensitive to the Majorana CP-violating phases 
$\alpha_{21}$, $\alpha_{31}$ \cite{BHP80,Lang86} -
information about these phases can be obtained
in the \betabeta-decay experiments 
\cite{BGKP96,BPP1,WR00,PPW,WR0203,PPR1} 
(see also  \cite{bb0nuCP1}).
Majorana CP-violating phases, 
and in particular, the phases $\alpha_{21}$ and/or 
$\alpha_{31}$, might be at the origin of 
the baryon asymmetry of the Universe \cite{LeptoG}. 

 One can express 
\cite{SPAS94} (see also, e.g., \cite{SPWIN99,BGGKP99,BPP1})
the masses $m_{2,3}$ 
and the elements
of the lepton mixing matrix entering into 
eq.\ (\ref{effmass2}) for \meff, in terms of 
the neutrino oscillation parameters measured in the 
solar and atmospheric neutrino and KamLAND
experiments: $m_{2,3}$  
--- in terms of the 
neutrino mass squared differences
\deltasol{} and \dma$\!\!$, driving the solar and 
atmospheric neutrino oscillations, 
and the mass $m_1$, and $|U_{\mathrm{e} j}|^2$, 
$j=1,2,3$, --- in terms of
the mixing angle which controls the solar 
$\nu_e$ transitions $\theta_{\odot}$,
and of the lepton mixing parameter $\sin^2\theta$
limited by the data from the CHOOZ
and Palo Verde experiments \cite{CHOOZ,PaloV}.

 The observation of \betabeta-decay
will have fundamental implications 
for our understanding of the symmetries of the
elementary particle interactions 
\footnote{Evidences for \betabeta-decay
taking place with a rate corresponding to
$0.11 \ {\rm eV} \leq  \meff  \leq  0.56$ eV
(95\% C.L.) are claimed to 
have been obtained in \cite{Klap01}. The
results announced in \cite{Klap01} have been 
criticized in \cite{bb0nu02}.}
(see, e.g., \cite{BiPet87}).
Under the general and plausible 
assumptions of 3-$\nu$ mixing, 
neutrino oscillation explanation 
of the solar and atmospheric neutrino 
data, massive Majorana neutrinos and
\betabeta-decay generated 
only by the (V-A) charged current weak interaction 
via the exchange of the three Majorana neutrinos,
which will be assumed to hold throughout this study,
the observation of \betabeta-decay
can provide unique information on 
\cite{BGGKP99,BPP1,PPW,WR0203,bb0nuMassSpec1,PPSNO2bb}
i) the type of neutrino mass spectrum which can be 
normal hierarchical (NH), inverted hierarchical (IH), or
quasi-degenerate (QD), 
ii) on the absolute scale of neutrino masses, 
and \cite{BGKP96,BPP1,WR00,PPW,WR0203,PPR1}
iii) on the Majorana CP-violating phases 
$\alpha_{21}$ and $\alpha_{31}$.

   A measured value of 
$\meff \sim {\rm few}\times 10^{-2}$ eV
can provide, in particular,
unique constraints on,
or even can allow one to determine, 
the type of the neutrino mass 
spectrum in the case 
$\nu_{1,2,3}$ are
Majorana particles
\cite{PPSNO2bb}.
The solar neutrino data 
and the first KamLAND results
\footnote{We assume throughout this study that
CPT-invariance holds in the lepton sector.},
favor relatively large value of 
$\cos2\theta_{\odot}$,
$\cos2\theta_{\odot} \sim 0.40$ 
\cite{fogli,valle,band,bahcall,others}.
A value of $\cos2\theta_{\odot} \gtap$ 0.25
would imply \cite{PPSNO2bb} 
the existence of significant lower 
bounds on \meff{} (exceeding 0.01 eV)
in the cases of IH 
and QD neutrino mass spectrum, 
and of a stringent upper bound 
(smaller than 0.01 eV) 
if the spectrum is of the NH type. The  
indicated lower bounds 
are in the range of the sensitivity of 
currently operating and planned 
\betabeta-decay experiments.

 Information on the absolute 
values of neutrino masses
in the range of interest
can also be derived in the 
\hbeta neutrino mass experiment KATRIN \cite{KATRIN},
and from cosmological and astrophysical 
data (see, e.g., ref.~\cite{Hu99}).

    Rather stringent upper
bounds on \meff{} have been obtained in the 
$^{76}$Ge experiments 
by the Heidelberg-Moscow collaboration \cite{76Ge00}, 
$ \meff  < 0.35~{\rm eV}$ ($90\%$ C.L.), 
and by the IGEX collaboration \cite{IGEX00},
$\meff  < (0.33 - 1.35)~{\rm eV}$ ($90\%$ C.L.).
Taking into account a factor of 3 uncertainty
in the calculated value of the corresponding 
nuclear matrix element \cite{ElliotVogel02},
we get for the upper limit
found in \cite{76Ge00}:  $\meff  < 1.05$ eV.
Higher sensitivity to 
\meff{} is planned to be 
reached in several $\betabeta$-decay experiments
of a new generation. 
The NEMO3 experiment \cite{NEMO3}, 
which began to take data in July of 2002, 
and the cryogenics detector CUORICINO 
\cite{CUORE} to be operative in 2003,
are expected to reach a sensitivity to values of 
$\meff  \sim 0.2~$eV.
Up to an order of magnitude better sensitivity, 
i.e., to $\meff  \cong 2.7\times 10^{-2}$ eV,
$1.5\times 10^{-2}~$eV, $5.0\times 10^{-2}~$eV,
$2.5\times 10^{-2}$ eV and $3.6\times 10^{-2}$ eV
is planned to be achieved 
in the CUORE, 
GENIUS,
EXO,
MAJORANA
and MOON
experiments \cite{CUORE} ~\footnote{The quoted sensitivities 
correspond to values of the relevant nuclear matrix
elements  from ref.\ \cite{SMutoKK90}.}, 
respectively.

   In what regards the \hbeta experiments, 
the currently existing most stringent upper 
bounds on the electron (anti-)neutrino mass  
$m_{\bar{\nu}_e}$ were obtained in the
Troitzk~\cite{MoscowH3} and Mainz~\cite{Mainz} 
experiments and read
$m_{\bar{\nu}_e} < 2.2$ eV\@.
The KATRIN \hbeta experiment \cite{KATRIN}
is planned to reach a sensitivity  
to  $m_{\bar{\nu}_e} \sim 0.35$ eV.

  In the present article we study
in detail the possibility of
determining the type, 
or excluding one or more types,
of neutrino mass spectrum 
by measuring of \meff{}  
in the next generation of \betabeta-decay 
experiments. The three possible 
types of 
spectra are considered~
\footnote{We work with the convention
$m_1 < m_2 < m_3$ and 
use the term ``spectrum with normal (inverted) 
hierarchy'' for the spectra 
with $\deltasol \!\! \equiv \deltadue \!\!$
($\deltasol \equiv \deltatre \!\!$), 
while we call ``normal 
hierarchical (NH)'' (``inverted hierarchical (IH)'')
the neutrino mass spectrum with normal 
(inverted) hierarchy and $m_1 \ll m_2, m_3$.}:
i) hierarchical (NH) $m_1 \ll m_2 \ll m_3$,
ii) inverted hierarchical (IH),
$m_1 \ll m_2 \cong m_3$, and
iii) quasi-degenerate (QD),
$m_1 \cong m_2 \cong m_3 \equiv m_0$,
$m^2_{1,2,3} \gg \dma\!\!$. 
In our analysis we take into account, 
in particular, the uncertainty in 
the determination of \meff{} 
due to the imprecise knowledge of
the relevant nuclear matrix elements.
This permits us to determine
the requirements which 
the possibility of distinguishing between
i) the NH and IH, ii) the NH and QD, and
iii) the IH and QD spectra, imposes on the 
uncertainty in the values
of the \betabeta-decay nuclear 
matrix elements. We derive also the 
maximal values of $\tan^2\theta_{\odot}$
for which the measurement
of \meff{}  would allow one
to discriminate between
the NH and IH, NH and QD and IH and QD spectra,
for different given values
of the nuclear matrix element
uncertainty factor.
An upper limit $\meff < {\rm few}\times 10^{-2}$ eV
would imply 
a significant constraint on the type of 
neutrino mass spectrum in the case the 
massive neutrinos are Majorana particles,
e.g., on the theories in which the
neutrino masses are generated 
via the see-saw mechanism.

 It should be noted that 
the determination of the type of neutrino 
mass spectrum, based on the measured value of 
\meff, would provide simultaneously unique 
information 
on the absolute neutrino mass scale 
\cite{PPW,WR0203,bb0nuMassSpec1,PPSNO2bb}.
Similar information cannot be obtained 
in the neutrino oscillation experiments, 
in which the sign of \dma can be determined
\footnote{In the convention in which the sign of 
$\dma = \Delta m^2_{31}$ is not fixed, 
the latter determines the ordering
of the neutrino masses: $\dma \!\! >0$
corresponds to $m_1 < m_2 < m_3$, 
while $\dma \!\! <0$ implies
$m_3 < m_1 < m_2$.}
(see, e.g., \cite{AMMS,HLM}) since
neutrino oscillations depend on 
neutrino mass squared differences
and are insensitive
to the absolute neutrino mass scale.
The sign of \dma can be determined in
very long base-line neutrino oscillation 
experiments at neutrino factories
(see, e.g., \cite{AMMS}), and, e.g, using
combined data from long base-line
oscillation experiments at the JHF facility and
with off-axis neutrino beams \cite{HLM}.
Under certain rather special
conditions it might be determined also
in experiments with reactor
$\bar{\nu}_e$ \cite{SPMPiai01}. 

%%%%%%%%%%%%%%%%%%%%%%%%%%%%%%%%%%%%%%%%%%%%%%%%%%%%%%%%%%%%%%%
\vspace{-0.4cm}
\section{Neutrino Oscillation Data and the 
Effective Majorana Mass 
}
\vspace{-0.2cm}
%%%%%%%%%%%%%%%%%%%%%%%%%%%%%%%%%%%%%%%%%%%%%%%%%%%

\hskip 1.0cm  The predicted value 
of \meff{} depends in the 
case of $3-\nu$ mixing on:
i) \dma$\!\!$,
ii) $\theta_{\odot}$ and $\Delta m^2_{\odot}$, 
iii) the lightest neutrino mass, 
and on iv) the mixing angle $\theta$.
Using the convention $m_1 < m_2 < m_3$, 
one has $\dma \equiv \Delta m^2_{31}$,
where $\Delta m^2_{jk} \equiv m_j^2 - m_k^2$,
and $m_3 = \sqrt{m_1^2 + \dma}$, while
either $\deltasol \equiv \Delta m^2_{21}$ 
or $\deltasol \equiv \Delta m^2_{32}$.
The two possibilities for \deltasol{}
correspond respectively to the two different types
of neutrino mass spectrum ---  
with normal and with inverted hierarchy.
In the first case we have 
$m_2 = \sqrt{m_1^2 + \deltasol}$,
$|U_{\mathrm{e} 1}|^2 = \cos^2\theta_{\odot} (1 - |U_{\mathrm{e} 3}|^2)$, 
$|U_{\mathrm{e} 2}|^2 = \sin^2\theta_{\odot} (1 - |U_{\mathrm{e} 3}|^2)$,
and  $|U_{\mathrm{e} 3}|^2 \equiv \sin^2\theta$,
while in the second 
$m_2 = \sqrt{m_1^2 + \dma - \deltasol}$,
$|U_{\mathrm{e} 2}|^2 = \cos^2\theta_{\odot} (1 - |U_{\mathrm{e} 1}|^2)$, 
$|U_{\mathrm{e} 3}|^2 = \sin^2\theta_{\odot} (1 - |U_{\mathrm{e} 1}|^2)$,
and  $|U_{\mathrm{e} 1}|^2 \equiv \sin^2\theta$.

     Given \dms$\!\!$, \dma$\!\!$, $\theta_{\odot}$ and
$\sin^2\theta$, the value of \meff{} 
depends strongly on the type of the
neutrino mass spectrum, as well as 
on the values of the two
Majorana CP-violating phases,
$\alpha_{21}$ and $\alpha_{31}$ 
(see eq.\ (\ref{effmass2})),
present in the lepton mixing matrix.
Let us note that in the case 
of QD spectrum,
$m_1 \cong m_2 \cong m_3$, 
$m_{1,2,3}^2 \gg \dma\!\!, \deltasol$,
~\meff{} is essentially independent on
\dma and \deltasol, and 
the two possibilities, 
$\deltasol \equiv \deltadue$
and $\deltasol \equiv \deltatre \!\!$, 
lead {\it effectively} 
to the same predictions for  
\meff{} 
\footnote{This statement is valid, 
within the convention 
$m_1 < m_2 < m_3$ we are using,
as long as there are no independent
constraints on the  CP-violating phases
$\alpha_{21}$ and $\alpha_{31}$ 
which enter into the expression for \meff.
In the case of NH spectrum, \meff{}  depends primarily 
on $\alpha_{21}$ ($|U_{\mathrm{e} 3}|^2 \ll 1$), 
while if the spectrum is with IH,
\meff{}  will depend essentially
on $\alpha_{31} - \alpha_{21}$ 
($|U_{\mathrm{e} 1}|^2 \ll 1$).}.

   The possibility of determining the type of the
neutrino mass spectrum
if \meff{} is found to be nonzero in the 
\betabeta-decay experiments of the next 
generation, depends crucially  
on the precision with which 
\dma$\!\!$, $\theta_{\odot}$, \deltasol, 
 $\sin^2\theta$ and \meff{} 
will be measured. It depends also crucially
on the values of
$\theta_{\odot}$ and of \meff.
The precision itself of the measurement of 
\meff{} in the next generation of \betabeta-
decay experiments, given the latter sensitivity
limits of $\sim (1.5 - 5.0)\times 10^{-2}~{\rm eV}$,
depends on the value of \meff.   

  The KATRIN experiment \cite{KATRIN}  
can test the hypothesis of a QD 
spectrum 
\footnote{Given the allowed regions of values of 
\deltasol{} and $\dma$ \cite{SKatm00}, 
one has a QD spectrum
for $m_{1,2,3}\cong m_{\bar{\nu}_e}>0.20$ eV.},
provided $m_{1,2,3} \cong m_{\bar{\nu}_e} \gtap (0.35 - 0.40)$ eV.
The KATRIN detector is designed to have a
1 s.d.\ error of 0.08 eV$^2$ on a measured value of
$m_{\bar{\nu}_e}^2$. 
This experiment is expected to start in 2007.

   Assuming CPT-invariance, combined 
$\nu_e \rightarrow \nu_{\mu (\tau)}$
and $\bar{\nu_e} \rightarrow \bar{\nu}_{\mu (\tau)}$
oscillation analyzes of the solar neutrino data
and of the just published 
first KamLAND results \cite{KamLAND},
have already been performed 
in \cite{fogli,valle,band,bahcall,others}.
All analyzes show
that the data favor the 
LMA MSW solution with $\deltasol > 0$
and $\tan^2\theta_{\odot} < 1$,
all the other solutions (LOW, VO, etc.) 
being essentially ruled out.
In Tables 1 and 2 we give the best-fit values 
and the 90\% C.L.\ allowed ranges
of $\Delta m^2_{\odot}$ and
$\tan^2\theta_{\odot}$
in the LMA solution region
obtained in \cite{fogli,valle,band,bahcall}.
The best fit values are confined to the narrow intervals
$(\Delta m^2_{\odot})_{\mathrm{BF}} = 
(6.9 - 7.3)\times 10^{-5}~{\rm eV^2}$, 
$(\tan^2\theta_{\odot})_{\mathrm{BF}} = (0.42 - 0.46)$.
The latter corresponds to 
$(\cos2\theta_{\odot})_{\mathrm{BF}} = (0.37 - 0.41)$.

  In the two-neutrino $\nu_{\mu} \rightarrow \nu_{\tau}$
($\bar{\nu}_{\mu} \rightarrow \bar{\nu}_{\tau}$)
oscillation analysis of the 
SK atmospheric neutrino
data performed in \cite{SKatm00} 
the following best-fit value of \dma  
was obtained: $(\dma \!\!)_{\mathrm{BF}} 
\cong 2.5\times 10^{-3}~{\rm eV^2}$.
At 99.73\% C.L., \dma was found to lie 
in the interval: $(1.5 - 5.0)\times 10^{-3}~{\rm eV^2}$. 
According to the more recent 
combined analysis of the data from 
the SK and K2K experiments \cite{fogliold},
one has $\dma \!\! 
\cong (2.7 \pm 0.4)\times 10^{-3}~{\rm eV^2}$. 
In certain cases of our analysis we will use 
as illustrative ``best-fit'' values 
$(\Delta m^2_{\odot})_{\mathrm{BF}} = 
7.0\times 10^{-5}~{\rm eV^2}$ and 
$(\dma\!\!)_{\mathrm{BF}} 
= 3.0\times 10^{-3}~{\rm eV^2}$.

   For the indicated allowed ranges of values of
$\Delta m^2_{\odot}$ and $\dma$, the NH (IH) spectrum
corresponds to $m_1 \ltap 10^{-3}~(2\times 10^{-2})$ eV. 

   A 3-$\nu$ oscillation analysis of the CHOOZ data 
showed \cite{BNPChooz} that 
for $\deltasol \ltap 10^{-4}~{\rm eV^2}$,
the limits on $\sin^2\theta$ practically coincide with
those derived in the 2-$\nu$ oscillation analysis
in \cite{CHOOZ}.
Combined 3-$\nu$ oscillation analysis of the 
solar neutrino, 
CHOOZ and the 
KamLAND data was performed
in \cite{fogli}
under the assumption of $\deltasol \ll \dma \!\!$
(see, e.g., \cite{BGG99,SPWIN99,ADE80}).
For the best-fit value of
$\sin^2\theta$ the authors of 
\cite{fogli} obtained:
$(\sin^2\theta)_{\mathrm{BF}} \cong (0.00 - 0.01)$. 
It was also found in \cite{fogli}
that $\sin^2\theta < 0.05$ at 99.73\% C.L. 

 The existing solar neutrino and KamLAND 
data favor values of 
$\Delta m^2_{\odot} \cong 
(5.0 - 10.0)\times 10^{-5}~{\rm eV^2}$ 
\cite{fogli,valle,band,bahcall,others}.
If $\Delta m^2_{\odot}$ lies in 
this interval, 
a combined analysis of the 
future more precise KamLAND 
results and of the solar neutrino data 
would permit to determine
the values of $\Delta m^2_{\odot}$ and
$\tan^2\theta_{\odot}$ with high precision: 
the estimated (1 s.d.) errors on 
$\Delta m^2_{\odot}$ and on
$\tan^2\theta_{\odot}$ 
can be as small as
$\sim (3 - 5)\%$ and
$\sim 5\%$ (see, e.g., \cite{Carlos01,fogliold}). 

   Similarly, if \dma lies in the interval 
$\dma \!\!\cong (2.0 - 5.0)\times 10^{-3}~{\rm eV^2}$, 
as is suggested by the current 
atmospheric neutrino data \cite{SKatm00,fogliold}, 
its value will be determined with a 
$\sim 10\%$ error (1 s.d.)  by the MINOS experiment \cite{MINOS}.
Somewhat better limits on $\sin^2 \theta$ than 
the existing one can be obtained in the 
MINOS experiment \cite{MINOS} as well. 
Various options are being currently discussed
(experiments with off-axis neutrino beams, more precise
reactor antineutrino and long base-line experiments, etc.,
see, e.g., \cite{MSpironu02}) of how to improve
by at least an order of magnitude, i.e., 
to values of $\sim 0.005$ or smaller, 
the sensitivity to $\sin^2\theta$. 

  The high precision measurements of 
\dma$\!\!$, $\tan^2\theta_{\odot}$ and
\deltasol~are expected to take place 
within the next $\sim (6 - 7)$ years. 
We will assume in what follows that 
the problem of measuring
or tightly constraining $\sin^2\theta$ will also be  
resolved  within the indicated period.
Under these conditions, 
the largest uncertainty
in the comparison of the theoretically 
predicted value of \meff{} with that 
determined in the
\betabeta-decay experiments would be associated
with the corresponding \betabeta-decay
nuclear matrix elements. 
We will also assume in what follows that
by the time one or more \betabeta-decay 
experiments of the next generation
will be operative ($2009 - 2010$)
at least the physical range
of variation of the values of the relevant
\betabeta-decay nuclear matrix elements 
will be unambiguously determined.

%%%%%%%%%%%%%%%%%%%%%%%%%%%%%%%%%%%%%%%%%%%%%%%%%
\vspace{-0.5cm}
\section{Determining the Type of Neutrino Mass Spectrum} 
\vspace{-0.2cm}
%%%%%%%%%%%%%%%%%%%%%%%%%%%%%%%%%%%%%%%%%%%%%%%%

\hskip 1.0cm 
   The possibility to distinguish
between the three different types of neutrino mass spectrum
in the 3-neutrino mixing case 
under discussion
depends on the allowed ranges of values of \meff{} for the 
three spectra. More specifically, it is determined by 
the maximal values of \meff{} in the cases of NH and IH spectra
and by the minimal values of \meff{} for the IH and QD spectra.
For the NH neutrino mass spectrum 
($m_1 \ll m_2 \ll m_3$), 
the maximal value of \meff{} is 
obtained in the case of CP-conservation 
and equal CP-parities of $\nu_{1,2,3}$:
%%%%%%%%%%%%%%%%%%%%%%%%%%%
\begin{equation} \label{eq:meffNH}
\meff_{\rm max}^{\rm NH} \cong \frac{1 - s^2}{1 + \ts} \left(m_1 +
\ts \, \sqrt{\dms} + (1 + \ts) \, \frac{s^2}{1 - s^2} \, \sqrt{\dma} \right)~,
\end{equation}
%%%%%%%%%%%%%%%%%%%%%%%%%%%%%%%
%
\noindent where $s^2 \equiv \sin^2\theta$
and we have neglected $m_1^2$ with respect to 
$\deltasol $ and \dma$\!\!$.

   In the case of IH neutrino mass spectrum 
($m_1 \ll m_2 \cong m_3$),
the effective Majorana mass lies in the
interval \cite{BGKP96,BPP1}
%%%%%%%%%%%%%%%%%%%%%%%%%%%%%%%
\begin{equation} \label{eq:meffIH}
\meff_{\rm min}^{\rm IH} \le \meff^{\rm IH} \le \meff_{\rm max}^{\rm IH}~,
\end{equation}
%%%%%%%%%%%%%%%%%%%%%%%%%%%%%%%%%
with 
%%%%%%%%%%%%%%%%%%%%%%%%%%%%%%%
\begin{equation} \label{eq:IHminmax}
\meff_{\rm min}^{\rm IH}  
\cong  
(1 - s^2) \, \cos 2 \theta_\odot 
\, \sqrt{\dma},~~~
\meff_{\rm max}^{\rm IH} 
\cong 
(1 - s^2) \, \sqrt{\dma}, 
\end{equation}
%%%%%%%%%%%%%%%%%%%%%%%%%%%%%%%
where we have neglected $m_1$.
The minimal (maximal) value of \meff, $\meff_{\rm min}^{\rm IH}$ 
($\meff_{\rm max}^{\rm IH}$), corresponds to CP-conservation
and opposite (equal) CP-parities of the neutrinos
$\nu_2$ and $\nu_3$.

   The minimal value of \meff{} for 
the quasi-degenerate (QD) 
neutrino mass spectrum ($m_1 \cong m_2 \cong m_3 \equiv m_0$, 
$m^2_0 \gg \dms \!\!, \dma\!\!$), for fixed value of $m_0$ is given by
%%%%%%%%%%%%%%%%%%%%%%%%%%%%%%%%%%%%%%%%%%
\begin{equation} \label{eq:meffQD}
\meff_{\rm min}^{\rm QD} \cong \frac{1 - s^2}{1 + \ts} \left( 1 - \ts
- \frac{s^2}{1 - s^2} (1 + \ts)\right) \, m_0~,
\end{equation}
%%%%%%%%%%%%%%%%%%%%%%%%
where $m_0 \gs 0.20$ eV\@ and we have neglected 
$\deltasol$ and \dma with respect to $m_0^2$.
As eq.~(\ref{eq:meffQD}) shows,
$\meff_{\rm min}^{\rm QD}$ scales to a good approximation 
with $m_0$. Correspondingly, the minimal allowed value of
\meff{} for the QD mass  spectrum is obtained for
$m_0 = 0.2$ eV.

 In Tables 1 and 2 we show the calculated 
i) maximal predicted value of \meff{} 
in the case of NH neutrino mass spectrum, 
ii) the minimal value of \meff{} 
for the IH spectrum,
and iii) the minimal value of \meff{} 
for the  QD spectrum ($m_0 = 0.2$ eV),
for the best-fit and
the 90\% C.L.\ 
allowed ranges of
values of $\tan^2\theta_{\odot}$ and \deltasol~
in the LMA solution region.
In Table 3 we give the same quantities,
$\meff_{\rm max}^{\rm NH}$,
$\meff_{\rm min}^{\rm IH}$
and $\meff_{\rm min}^{\rm QD}$,
calculated using the best-fit values
of the neutrino oscillation parameters,
including 1 s.d. (3 s.d.) uncertainties 
of 5\% (15\%) on $\tan^2\theta_{\odot}$
and $\deltasol$ and of 10\% (30\%) on \dma$\!\!$.

 The maximal predicted value of \meff{} for
the IH spectrum is given by $\meff_{\rm max}^{\rm IH} 
\cong \sqrt{(\dma\!\!)_{\rm max}}$.
For the best-fit value \cite{SKatm00,fogliold}
and the 99.73\% C.L.\ allowed range \cite{SKatm00}
of \dma we have, respectively,
$\meff_{\rm max}^{\rm IH} \cong 0.05~{\rm and}~0.07$ eV. 

  On the basis of the results shown in Tables $1 - 3$, 
we can conclude, in particular, 
that the NH spectrum could be ruled out if the 
measured value of \meff{} 
exceeds approximately $0.9 \times 10^{-2}$ eV,
where we have been rather conservative
in choosing the maximal value.
 
%%%%%%%%%%%%%%%%%%%%%%%%%%%%%%%%%%%%%%%%%%%%%%%
\vspace{-0.4cm}
\subsection{\label{sec:unc}Theoretical and Experimental 
Uncertainties in \meff}
\vspace{-0.2cm}
%%%%%%%%%%%%%%%%%%%%%%%%%%%%%%%%%%%%%%%%%%%%%%%

\hskip 1.0cm Following the notation in ref.\ \cite{PPR1}, 
we will parametrize the 
uncertainty in \meff{}
due to the  imprecise knowledge of the 
relevant nuclear matrix elements ---
we will use the term ``theoretical uncertainty'' 
for the latter --- through a parameter $\zeta$, 
$\zeta \geq 1$, defined as:
%%%%%%%%%%%%%%%%%%%%%%%%%%%%%
\begin{equation} \label{eq:zeta}
\meff = \zeta \Big( (\meff_{\rm exp})_{\mbox{}_{\rm MIN}} \pm \Delta
\Big)~,
\end{equation}%%%%%%%%%%%%%%%%%%%%%%%%%%%%%
where $(\meff_{\rm exp})_{\mbox{}_{\rm MIN}}$ 
is the value of \meff{}
obtained from the measured 
\betabeta-decay half life-time
of a given nucleus using 
{\it the largest nuclear matrix element}
and $\Delta$ is  the experimental error. 
An experiment measuring a \betabeta-decay 
half-life time 
will thus determine a range of \meff{} corresponding to 
%%%%%%%%%%%%%%%%%%%%%%%%%%
\begin{equation}
(\meff_{\rm exp})_{\mbox{}_{\rm MIN}} - \Delta 
\le \meff  \le 
\zeta \Big( (\meff_{\rm exp})_{\mbox{}_{\rm MIN}} + \Delta \Big)~. 
\end{equation}
%%%%%%%%%%%%%%%%%%%%%%%%%%%%%%%%%
The currently estimated range of 
$\zeta$ for experimentally interesting 
nuclei varies from 3.5 for 
$^{48}$Ca to 38.7 for $^{130}$Te, 
see, e.g., Table 2 in ref.\ \cite{ElliotVogel02}
and ref.\ \cite{bilgri}. 

   We estimate, following again \cite{PPR1}, the 1 s.d.\ 
error on the experimentally 
measured value of \meff{} by using the standard 
expression
%%%%%%%%%%%%%%%%%%%%%%%%%
\begin{equation}
  \frac{\sigma (\meff)}{\meff} = \sqrt{ (E_1)^2 + (E_2)^2}~,
\label{eq:sigma}
\end{equation}
%%%%%%%%%%%%%%%%%%%%%%%%
where $E_1$ and $E_2$ are the statistical and systematic
errors. We choose $E_2 = const = 0.05$ and 
take $E_1 = f/\meff$, where we assume $f$ = 0.028 eV.
This gives a total relative error 
$\sigma (\meff)/\meff \cong 15\%$ at $\meff = 0.20$ eV.
The above choices were motivated by the fact that
the sensitivities of the next generation of 
\betabeta-decay experiments 
in the measurement of \meff{} 
are estimated to be in the range
of $\sim (1.5 - 5.0)\times 10^{-2}$ eV and if, e.g,
$\meff \gtap 0.20$ eV, a precision in the determination
of \meff{} corresponding to an error of $\sim 15\%$
could be reached in these experiments. 
Moreover, for values of \meff{} which are 
sufficiently bigger than the quoted 
sensitivity limits of the
future experiments, the statistical error
scales as \meff{} increases like $E_1 \sim const/\meff$.

%%%%%%%%%%%%%%%%%%%%%%%%%%%%%%%%%
\vspace{-0.4cm}
\subsection{Requirements on the Solar Neutrino Mixing Angle}
\vspace{-0.2cm}
%%%%%%%%%%%%%%%%%%%%%%%%%%%%%%%%%%

\hskip 1.0truecm We shall derive next 
the constraints $\ts$ must satisfy
in order to be possible to distinguish 
between the three types of neutrino mass spectrum
NH, IH and QD. 
%%%%%%%%%%%%%%%%%%%%%%%%%%%

\vspace{0.2cm}
\noindent {\bf Case i). Normal Hierarchical 
and Inverted Hierarchical Spectra}.
In order to be possible to 
distinguish between 
the NH and IH spectra, 
the following inequality 
must hold:
%%%%%%%%%%%%%%%%%%%%%%
\begin{equation} \label{eq:NHIHcond}
\zeta \, \meff_{\rm max}^{\rm NH} < \meff_{\rm min}^{\rm IH}~,~\zeta \ge 1~. 
\end{equation}
%%%%%%%%%%%%%%%%%%%%%%
%
From this inequality, 
using eqs.\ (\ref{eq:meffNH}) and (\ref{eq:IHminmax}), 
we get:
%%%%%%%%%%%%%%%%%%%%%%%%%%%%%%%
\begin{equation} \label{eq:tsNHIH}
\ts  < \frac{1 - \zeta \, (\beta + t^2)}{1 + \zeta \, (\alpha + t^2)}~,
\end{equation}
%%%%%%%%%%%%%%%%%%%%%%%%%%%%%%%
%
where $t^2 = s^2/(1 - s^2)$,
$\alpha = \sqrt{\dms/\dma}$ and
$\beta = \sqrt{m_1^2/\dma}$.
For our illustrative ``best-fit'' values
$(\Delta m^2_{\odot})_{\mathrm{BF}} = 
7.0\times 10^{-5}~{\rm eV^2}$ and 
$(\dma\!\!)_{\mathrm{BF}} 
\cong 3.0\times 10^{-3}~{\rm eV^2}$,
one has $\alpha \simeq 0.153$; with
$m_1 \ls 0.001$ eV one also finds $\beta \ls 0.018$. 
For $\zeta = 1$, the indicated values of
$\alpha$ and $\beta$ and $s^2 = 0.05~(0)$, 
eq.\ (\ref{eq:tsNHIH})
is fulfilled for $\ts \ls 0.77~(0.85)$.
Taking $\zeta = 2$, one finds 
$\ts \ls 0.61~(0.74)$,
while for $\zeta = 3$ the result is 
$\ts \ls 0.49~(0.65)$.

The smaller $m_1$ and/or \dms$\!\!$, 
the closer the upper bound on \ts{}
of interest becomes to 1.
The above analysis shows also that
the upper bound
on $\ts$ under discussion
exhibits relatively strong dependence
on the value of $s^2 \ls 0.05$:
it increases by a factor of $\sim (1.2 - 1.5)$
when $s^2$ decreases from 0.05 to 0.  

%%%%%%%%%%%%%%%%%%%%%%%%
% \item[ii)] 
\vspace{0.2cm}
\noindent {\bf Case ii). Normal Hierarchical  
and Quasi-Degenerate Spectra. 
}
Distinguishing between the NH and QD
spectra requires that the following inequality is 
satisfied:
%%%%%%%%%%%%%%%%%%%%%%%%%%%%%%%
\begin{equation} 
\zeta \, \meff_{\rm max}^{\rm NH} < \meff_{\rm min}^{\rm QD}~,~\zeta \ge 1~.
\end{equation}
%%%%%%%%%%%%%%%%%%%%%%%%%%%%%%%
%
From this inequality using eqs.\ (\ref{eq:meffNH}) and 
(\ref{eq:meffQD}) we get:
%%%%%%%%%%%%%%%%%%%%%%%%%%%%%%%%%
\begin{equation} \label{eq:tsNHQD}
\ts < \frac{1 - \zeta \, \tilde{\beta} - t^2 \, 
(1 + \zeta \, \tilde{\gamma})}
{1 + \zeta \, \tilde{\alpha} + t^2 \, (1 + \zeta \, \tilde{\gamma})}~,
\end{equation}
%%%%%%%%%%%%%%%%%%%%%%%%%%%%%%%%%
%
where $\tilde{\alpha} = \sqrt{\dms/m_0^2}$,
$\tilde{\beta} = m_1/m_0$ and 
$\tilde{\gamma} =
\sqrt{\dma/m_0^2}$.
For our illustrative ``best-fit'' 
values of $\deltasol$ and \dma
and $m_0 \gs 0.2$ eV, we have:
$\tilde{\alpha} \ls 0.042$,
$\tilde{\beta} \ls 0.000025$, and
$\tilde{\gamma} \ls 0.274$. 
Using these upper limits we find that for 
$s^2 = 0.05~(0)$ and $\zeta = 1$,
eq.\ (\ref{eq:tsNHQD}) is satisfied
if $\ts \ls 0.9~(1.0)$. 
Taking $\zeta = 2,~3$ 
we get $\ts \ls 0.8~(1.0)$ for $s^2 = 0.05~(0)$. 
If, e.g., $m_0 = 2.0$ eV, one finds 
$\tilde{\gamma} \ls 0.035$, i.e., 
the larger the value of 
$m_0$, the smaller 
$\tilde{\gamma}$ and 
the closer is the upper bound on $\ts$ to 1,
i.e., the less constraining it is. 
Since $\tilde{\gamma}$ enters into eq.\ (\ref{eq:tsNHQD}) 
multiplied by the relatively
small quantity $t^2$, the deviation of 
the upper bound on \ts{} under discussion 
from 1 is determined essentially by the value of
$\tilde{\alpha}$. Correspondingly, 
the maximal value of \ts{} 
permitting to distinguish between
the NH and QD neutrino mass spectra
decreases with increasing of \dms$\!\!$. 

\vspace{0.2cm}
\noindent {\bf Case iii). Inverted Hierarchical 
and Quasi-Degenerate Spectrum.} 
One could distinguish between these 
two types of spectra
if the following inequality
is fulfilled:
%%%%%%%%%%%%%%%%%%%%%%%%%%%%%%%%%%%%
\begin{equation} 
\zeta \, \meff_{\rm max}^{\rm IH} 
< \meff_{\rm min}^{\rm QD}~,~\zeta \ge 1~.
\label{eq:IHQDcond}
\end{equation}
%%%%%%%%%%%%%%%%%%%%%%%%%%%%%%%%%%%%
%
This condition together with eqs.\ (\ref{eq:IHminmax}) and 
(\ref{eq:meffQD}) leads to the constraint
%%%%%%%%%%%%%%%%%%%%%%%%%%%%%%%%%
\begin{equation} \label{eq:tsIHQD}
\ts < \frac{1 - \zeta \, \tilde{\gamma} - t^2} {1 + \zeta \, 
\tilde{\gamma} + t^2}~,
\end{equation}
%%%%%%%%%%%%%%%%%%%%%%%%%%%%%%%%%%%%%
%
where $\tilde{\gamma}$ was defined earlier.  
Using our illustrative
$\deltasol$ and \dma
``best-fit'' values and $m_0 \gs 0.2$ eV,
one finds $\tilde{\gamma} \ls 0.274$.
For $s^2 = 0.05~(0)$ and $\zeta = 1$, 
the above limit on  $\tilde{\gamma}$ 
together with 
eq.\ (\ref{eq:tsIHQD}) leads to
$\ts \ls 0.5~(0.6)$. 
Larger values of $\zeta$ lead to 
stringent restrictions  
on \ts: for $\zeta = 2$, for instance,
we find $\ts \ls 0.2~(0.3)$ 
for $s^2 = 0.05~(0)$. 
The requirement that the two  
spectra could be distinguished 
is less restrictive 
for larger values of $m_0$  
in this case as well. 

\vspace{2mm}
  These simple quantitative analyses show 
that if \meff{} is found to be non-zero
in the future \betabeta-decay experiments,
it would be easier, in general, to distinguish 
between the spectrum with NH
and those with IH or of QD type
using the data on $\meff \neq 0$,
than to distinguish between the 
IH and the QD 
spectra. Discriminating between the 
latter would be less 
demanding if $m_0$ is sufficiently large.
The requirement of distinguishing  between the NH
and the QD spectra leads to  
the least stringent conditions.  

   The above analyses also show that
the possibility to distinguish between 
the IH and QD, and NH and QD, spectra
depends rather weakly on the value $s^2$,
satisfying the existing upper limits 
\cite{CHOOZ,PaloV,fogliold}:
the relevant upper bounds on 
\ts{} decrease somewhat 
with decreasing of $s^2$. 
This is not so in the case
of NH versus IH spectra: 
the upper bound of interest can increase
noticeably (e.g., by a 
factor of $\sim (1.2 - 1.5)$)
when $s^2$ decreases from 0.05 to 0.
  
   It is worth noting 
that in contrast to the conditions
which would allow one to establish
on the basis of a measurement of
$\meff \neq 0$ the 
presence of CP violation 
due to the Majorana CP-violating 
phases \cite{PPR1}, 
the conditions permitting to distinguish between
the three types of neutrino mass spectrum
imply an {\it upper limit} on $\ts$.

\begin{figure}
\begin{center}\hspace{-2cm}
\epsfig{file=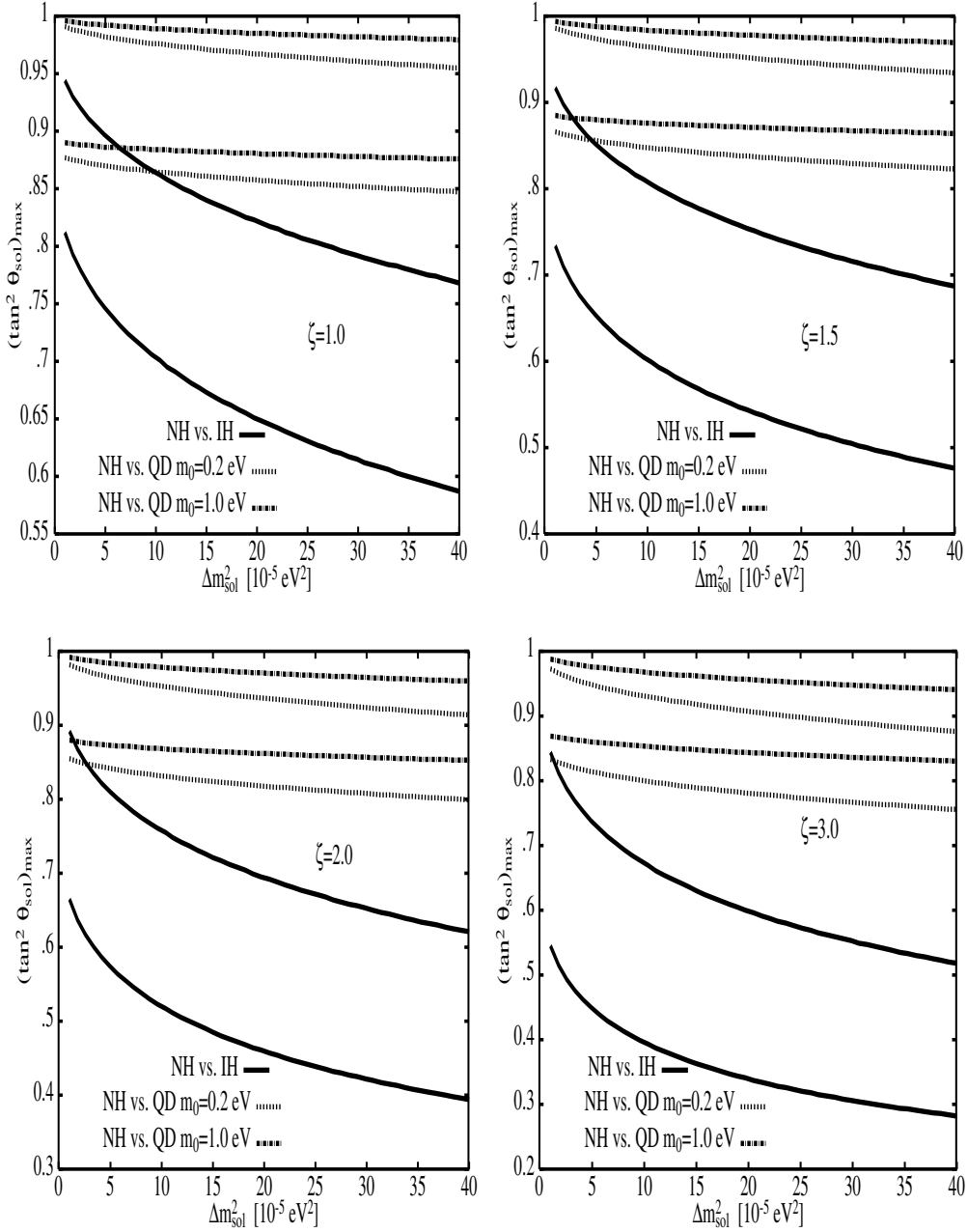,width=15cm,height=20cm}
\caption{\label{fig:spr1} The upper bound on \ts{}, 
for which one 
can distinguish the NH
spectrum from the IH spectrum 
and from that of 
QD type, as a function of \dms$\!\!$
for $\dma\!\!= 3 \times 10^{-3} \eV^2$
and different values of $\zeta$
(see eqs.\ (\ref{eq:tsNHIH}) and (\ref{eq:tsNHQD})). 
The lower (upper) line corresponds to $s^2 = 0.05$ (0).
}
\end{center}
\end{figure}

    In Fig.\ \ref{fig:spr1} we show 
the upper bounds on \ts{}, for which one 
can distinguish the NH
spectrum from the IH spectrum 
and from that of 
QD type, as a function of \dms$\!\!$
for different values of $\zeta$. 
As is seen from the figure,
the dependence of the 
maximal value of \ts{} of interest
on $m_0$ in both cases is  modest. 
Obviously, with the increasing 
of \dms and/or $s^2$, 
$\meff_{\rm max}^{\rm NH}$  
also increases. As a consequence,
the maximal \ts{} under discussion decreases, 
which means that the corresponding spectra
become harder to distinguish.

\begin{figure}
\begin{center}\hspace{-2cm}
\epsfig{file=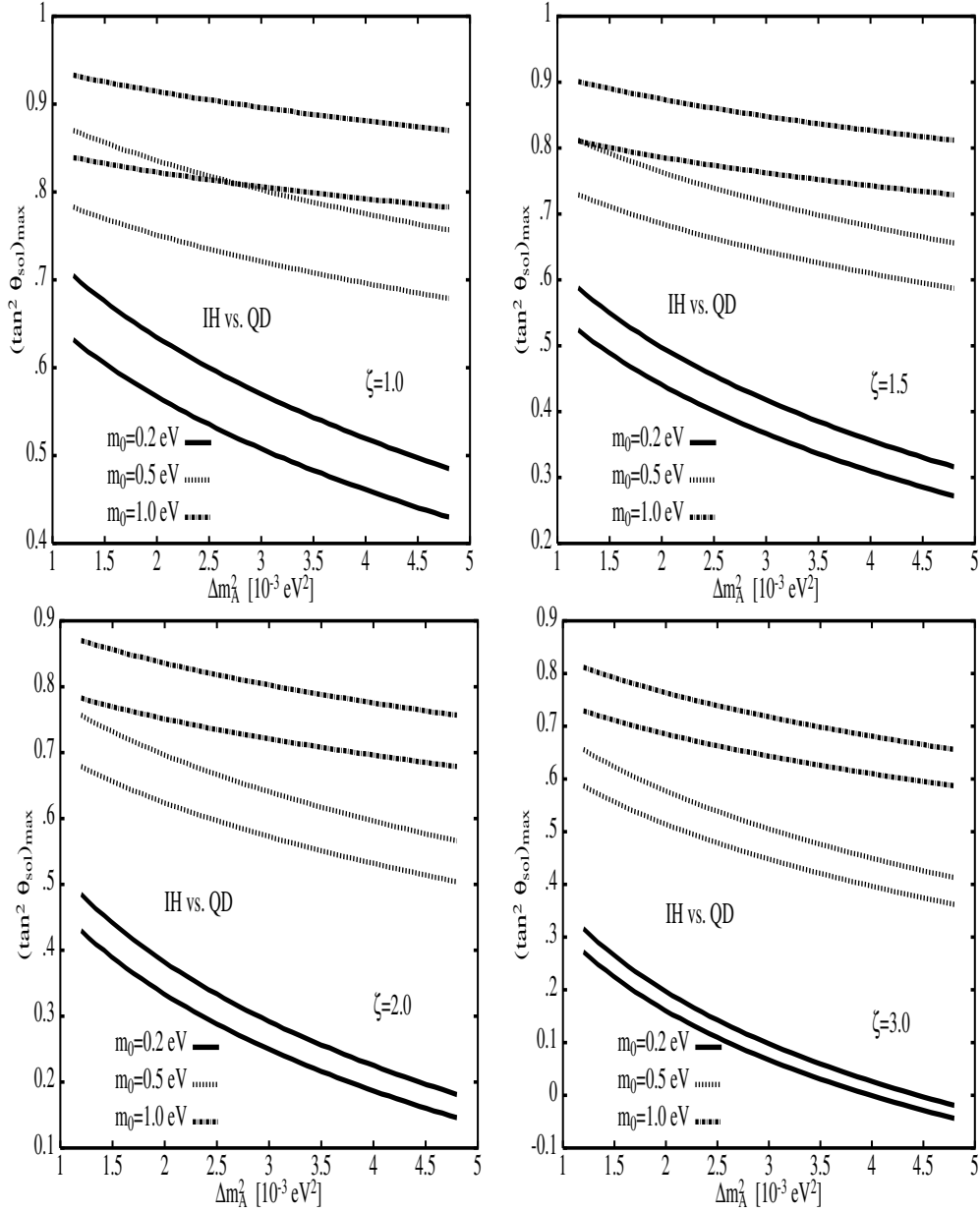,width=15cm,height=20cm}
\caption{\label{fig:spr2} The upper bound on \ts{} 
allowing one to discriminate between 
the IH and the QD neutrino mass spectra,
as a function of \dma for different 
values of $\zeta$ (see eq.\ (\ref{eq:tsIHQD})). 
The lower (upper) line corresponds to $s^2 = 0.05$ (0).}
\end{center}
\end{figure}

 As we have seen, 
in order to be possible
to distinguish between the IH and the QD spectra
eq.\ (\ref{eq:tsIHQD}) should be fulfilled.
Fig.\ \ref{fig:spr2} shows 
the upper bound on \ts{} as implied by 
eq.\ (\ref{eq:tsIHQD}), 
for $s^2 = 0.05~{\rm and}~ 0.0$
as a function of \dma$\!\!$. 
The upper bound on $\ts$ 
of interest depends strongly on the value of
$m_0$. It decreases with the increasing
of \dma$\!\!$, the dependence on
\dma being noticeable for
$m_0 \cong 0.20$ eV and
rather mild for 
$m_0 \gtap 0.40$ eV.
As it follows from Fig.\ \ref{fig:spr2},
for the values of 
\dma favored by the neutrino oscillation
data and for $\zeta \gtap 2$, distinguishing
between the IH and QD spectra in the
case of $m_0 \cong 0.20$ eV requires 
too small,
from the point of view of the existing data,
values  of $\ts$. For
$m_0 \gtap 0.40$ eV, the values of $\ts$ 
of interest fall in the ranges
favored by the existing solar neutrino 
and KamLAND data
even for $\zeta = 3$.

%%%%%%%%%%%%%%%%%%%%%%%%%%%%%
\vspace{-0.4cm}
\subsection{\label{sec:exp} Requirements on $\Delta$ and $\zeta$
}
\vspace{-0.2cm}
%%%%%%%%%%%%%%%%%%%%%%%%%%%%%%%

\hskip 1.0cm We will investigate 
now the requirements the experimental 
and theoretical uncertainties
$\Delta$ and $\zeta$
should satisfy in order to allow one 
to discriminate between the three
different neutrino mass spectra 
if \meff{} is measured, 
or a significantly improved bound on \meff{}
is obtained.

%%%%%%%%%%%%%%%%%%%%%%%%%%%%%%%%%%%%%%%%%%
\vspace{-0.4cm}
\subsubsection{\label{sec:outQD} Testing  
the Quasi-Degenerate Neutrino Mass Spectrum} 
\vspace{-0.2cm}
%%%%%%%%%%%%%%%%%%%%%%%%%%%%%%%%%%%%%%%%

\hskip 1.0cm In order to rule out the QD spectrum 
it is necessary that
%%%%%%%%%%%%%%%
\begin{equation}
\zeta \left( \meffexp + \Delta \right) < \meff_{\rm min}^{ \mathrm{QD}}~,
\end{equation}
%%%%%%%%%%%%%%%
%
\vspace{-0.2cm}
which translates into a 
condition on the nuclear matrix element
uncertainty $\zeta$
%%%%%%%%%%%%%%%
\begin{equation}
\zeta < \frac{\meff_{\rm min}^{ \mathrm{QD}}}{\meffexp} 
 \left( 1 + \frac{\sigma(\meff)}{\meff} \right)^{-1}.
\label{eq:outQD1}
\end{equation}
%%%%%%%%%%%%%%%
%
For the illustrative 
sensitivities 
of the future \betabeta-decay experiments,
$\meffexp = 0.01;~0.02$; $0.04 \eV$,
negligible $\sigma(\meff)/\meff$,
and the predicted values of 
$\meff_{\rm min}^{ \mathrm{QD}}= (0.048-0.056) \eV$
reported in Table 3 (the 3 s.d. case), 
we have $\zeta < (4.8-5.6);~(2.4-2.8);~(1.2-1.4)$.
The better the sensitivity of the future experiments,
the larger is the allowed nuclear matrix element 
uncertainty. Including a non-negligible
$\sigma(\meff)/\meff$ makes 
even more restrictive
the condition on $\zeta$.

Proving that 
the neutrino mass spectrum 
is of the QD type requires that 
$(\meff_{\rm exp})_{\mbox{}_{\rm MIN}}-\Delta > \meff_{\rm min}^{\rm QD}$,
which implies an upper bound on $\Delta$:
%%%%%%%%%%%%%%%%%%%%%%%%%
\begin{equation} 
\Delta < (\meff_{\rm exp})_{\mbox{}_{\rm MIN}}- \meff_{\rm min}^{\rm QD}~. 
\end{equation}
%%%%%%%%%%%%%%%%%%%%%%%%%%%%%%%%%
%
Using the values of $\meff_{\rm min}^{\rm QD}$ 
reported in Table 3 (the 3 s.d. case), we find
the corresponding upper bounds
$\Delta < (52,52,44,46) \times 10^{-3} \ \eV$,
for $(\meff_{\rm exp})_{\mbox{}_{\rm MIN}}=0.1 \ \eV$,
and  $\Delta < (152,152,144,146) \times 10^{-3} \ \eV$,
for $(\meff_{\rm exp})_{\mbox{}_{\rm MIN}}=0.2 \ \eV$.
If the above condition is fulfilled,
condition~(\ref{eq:IHQDcond}) with $\zeta = 1$, which 
guarantees that the QD spectrum
is distinguishable from the NH and IH ones,
should also be satisfied.
For the illustrative ``best-fit'' values
$ \deltasol=7.0 \times 10^{-5} \ {\rm eV}^2$ and
$\dma = 3.0 \times 10^{-3} \ {\rm eV}^2$, 
and for $s^2 = 0.05 \, (0)$, eq.~(\ref{eq:IHQDcond}) 
($\zeta = 1$) holds
if $\tan^2 \theta_\odot < 0.5 \, (0.6)$.

%%%%%%%%%%%%%%%%%%%%%%%%%%%%%%%%%%
\vspace{-0.4cm}
\subsubsection{\label{sec:ruleoutNH} Testing the 
 Normal Hierarchical Spectrum 
} 
\vspace{-0.2cm}
%%%%%%%%%%%%%%%%%%%%%%%%%%%%%%%%%%%%%%%%%%%%

\hskip 1.0cm  The NH neutrino mass spectrum would be ruled out if 
%%%%%%%%%%%%%%%%%%%%%%%%%%%
\begin{equation}
(\meff_{\rm exp})_{\mbox{}_{\rm MIN}} - \Delta > \meff_{\rm max}^{\rm NH}~.  
\label{NHout1}
\end{equation}
%%%%%%%%%%%%%%%%%%%%
%
Parametrizing $(\meff_{\rm exp})_{\mbox{}_{\rm MIN}}$ as 
$(\meff_{\rm exp})_{\mbox{}_{\rm MIN}} = y^{\mbox{}_{\rm NH}}~
\meff_{\rm max}^{\rm NH}$, $y^{\mbox{}_{\rm NH}} > 1$,
we get
%%%%%%%%%%%%%%%%%%%%%%%%%%%%
\begin{equation}
\Delta < (y^{\mbox{}_{\rm NH}} - 1) \, \meff_{\rm max}^{\rm NH}~.
\label{NHout2}
\end{equation}
%%%%%%%%%%%%%%%%%%%%%%%%%%%%%%%%%%
%
 How restrictive this condition is 
depends on the value
of  $\meff_{\rm max}^{\rm NH}$. 
Assuming that the more precise measurements 
of $\tan^2\theta_{\odot}$, \dma
and $\sin^2\theta$ will not produce results
very different from their current best-fit values,
we can use the predictions for 
$\meff_{\rm max}^{\rm NH}$ given in Table 3 
(3 s.d. case):
$\meff_{\rm max}^{\rm NH}$ $\cong 0.0066$ eV.
With this value 
one finds that
for $y^{\mbox{}_{\rm NH}} = 40,~30,~20,~10,~7,~5$, 
condition (\ref{NHout2}) is satisfied if
$\Delta < (25.7,~19.1,~12.5,~5.9,~4.0,~2.6) \times 10^{-2}$ eV\@. 
Alternatively, if experimentally 
$\Delta = (0.30,~0.20,~0.10,~0.05)$ eV, 
condition (\ref{NHout2}) will hold
provided $y^{\mbox{}_{\rm NH}} > 46.5,~31.3,~16.2,~8.6$.

%%%%%%%%%%%%%%%%%%%%%%%%%%%%%%%%%%%%%%%%%%
\vspace{-0.4cm}
\subsubsection{\label{sec:outIH} Probing the Inverted 
Hierarchical Spectrum} 
\vspace{-0.2cm}
%%%%%%%%%%%%%%%%%%%%%%%%%%%%%%%%%%%%%%%%

\hskip 1.0cm  For the IH neutrino mass spectrum, $\meffih \ $ 
is constrained to lie in the interval given by 
eqs. (\ref{eq:meffIH}) and (\ref{eq:IHminmax}).
The IH spectrum can be ruled out if 
the experimentally measured value of \meff, 
with both the experimental error $\Delta$ and the
nuclear matrix element uncertainty factor $\zeta$
taken into account,
lies outside the range given in eq.~(\ref{eq:meffIH}).
There are two possibilities.

\vspace{0.2cm}
\noindent {\bf Case i).} 
%%%%%%%%%%%%%%%%%%%%%%%%%%%%%%%%%%%%%%%%%%%%%%%%%
\begin{equation}
\meffexp - \Delta > \meff_{\rm max}^{\rm IH}~,  
\label{eq:outIH3}
\end{equation} 
%%%%%%%%%%%%%%%%%%%%%%
where $\meff^{\rm IH}_{\rm max}$ depends on the allowed values of
\dma and $s^2$ and is given in the captions of Tables $1-3$.
Using  the parametrization
$(\meff_{\rm exp})_{\mbox{}_{\rm MIN}} = 
y^{\mbox{}^{\rm IH}}~\meff_{\rm max}^{\rm IH}$,
$y^{\mbox{}^{\rm IH}} > 1$, 
we are lead to the condition
%%%%%%%%%%%%%%%%%%%%%%%%%%%%%%%%%
\begin{equation} \label{eq:yIH}
\Delta < (y^{\rm IH} - 1) \, (1 - s^2) \, \sqrt{(\dma\!\!)_{\rm max}}~. 
\end{equation}
%%%%%%%%%%%%%%%%%%%%%%%%%%%%%%%%%
%
For $y^{\rm IH} = 5, 4, 3, 2, 1.5$ this condition 
is fulfilled if 
$\Delta < (0.28, 0.21, 0.14, 0.07, 0.03)$ ($1 - s^2$) eV\@. 
The larger the measured value of \meff, 
the larger is the maximal experimental 
error which still permits to rule out 
the IH spectrum. Alternatively, 
for a value of the experimental error  
$\Delta = (0.2, 0.1, 0.05, 0.03)$ eV 
it would be possible to rule out the
IH spectrum provided
$y^{\rm IH} > 3.9, 2.4, 1.7, 1.4$, respectively.

\vspace{0.2cm}
\noindent {\bf Case ii).}
The spectra with inverted hierarchy  can be ruled out also if:
%%%%%%%%%
\begin{equation}
\zeta \left( \meffexp + \Delta \right) 
< \meff_{\rm min}^{\rm IH}  \simeq
(1 -s^2) \, (\cos 2 \theta_\odot)_{\rm min}
\, \sqrt{\dma}~.
\label{eq:outIH2}
\end{equation}
%%%%%%%%%%%%%%%%%%%%%%
%
Since $\meff_{\rm min}^{\rm IH}$ is of 
the order of 0.01 eV, the experimental 
uncertainty will be  required to be 
even  below this value, making it
not within reach of the currently planed 
experiments, except possibly for the
10t version of GENIUS.
For instance, 
for $\meffexp \!\! = 0.01 \eV$ and $\Delta = 0.01 \eV$ one finds, e.g.,  
$\zeta < 1.1$ if $\meff_{\rm min}^{\rm IH} = 0.022\eV$. 

 Probing the IH neutrino mass spectrum requires that
the following conditions be fulfilled:
\begin{equation}
\label{eq:outIH4ab}
\meffexp - \Delta \geq \meff^{\mathrm{IH}}_{\mathrm{min}},~~~~
\zeta \left( \meffexp + \Delta \right) 
\leq \meff_{\rm max}^{\rm IH} ~.
\end{equation} 
%%%%%%%%%%%%%%%%%%%%%%
%
 Using the fact that 
$\meff_{\rm min}^{\rm IH}  = \cos 2 \theta_\odot \, 
\meff_{\rm max}^{\rm IH} $
and the parametrization 
$ \meffexp \!\! \equiv y^{\mbox{}^{\rm IH}} \meff_{\rm max}^{\rm IH} $, 
the necessary conditions on $\zeta$ and $y^{\mbox{}^{\rm IH}}$ read
%%%%%%%%%%%%%%%%%%%%%%%%%%%%%%%%%%%%%%%%%%%%%%%%%
\begin{equation}
y^{\mbox{}^{\rm IH}} \geq  \cos 2 \theta_\odot \, 
\left( 1 - \frac{\sigma(\meff)}{\meff} 
\right)^{-1},~~~~ 
\zeta \, y^{\mbox{}^{\rm IH}} \leq  \left( 1 + \frac{\sigma(\meff)}{\meff} \right)^{-1}.
\label{eq:outIH5}
\end{equation} 
%%%%%%%%%%%%%%%%%%%%%%
%

   In the most favorable situation in which 
$\sigma(\meff)/\meff \ll 1$
and $y^{\mbox{}^{\rm IH}}= \cos 2 \theta_\odot$, 
$\zeta$ is required to be
$\zeta < 1/\cos 2 \theta_\odot$. 
For the present best-fit values
of $\tan^2 \theta_\odot$ reported in Table 1, we obtain
$\zeta < 2.7, 2.7, 2.4, 2.5$. Let us note, however, that 
from experimental point of view this possibility is rather
demanding: as a first approximation, $\Delta$ has to be of the 
order of, or smaller than, the difference between the maximal 
and minimal values of \meff{} in the IH case. This difference 
is typically of the order of $\sim (0.02-0.04)$ eV, and
does not exceed $\sim 0.06$ eV. 

If conditions~(\ref{eq:outIH5}) are satisfied, 
in order to establish the IH spectrum
both eqs.~(\ref{eq:NHIHcond}) and (\ref{eq:IHQDcond}) 
with $\zeta=1$ should also be valid. 
Taking as illustrative values 
$ \deltasol=7.0 \times 10^{-5} \ {\rm eV}^2$ and
$\dma \!\! = 3.0 \times 10^{-3} \ {\rm eV}^2$, 
both conditions are satisfied for 
$\tan^2 \theta_\odot \ls 0.5 \ (0.6)$ if $s^2 = 0.05 \ (0)$.

%%%%%%%%%%%%%%%%%%%%%%%%%%%%%%%%%%%%%%%%%%
\vspace{-0.4cm}
\subsubsection{\label{sec:distIHQD} The Inverted Hierarchical versus
the Quasi-Degenerate 
Spectrum} 
\vspace{-0.2cm}
%%%%%%%%%%%%%%%%%%%%%%%%%%%%%%%%%%%%%%%%

\hskip 1.0cm 
Let us assume that a value 
of $(\meff_{\rm exp})_{\mbox{}_{\rm MIN}}$
of a few 10 meV has been found, 
thus ruling out the NH spectrum. The remaining 
question to ask in this situation would be 
whether the neutrino mass spectrum is of the  
IH or QD type. 
Distinguishing between the two types of 
spectra might be possible
provided 
%%%%%%%%%%%%%%%%%%%%%%%%%%%%%%%
\begin{equation} 
\meff_{\rm max}^{\rm IH} < \meff_{\rm min}^{\rm QD}~.
\end{equation}
%%%%%%%%%%%%%%%%%%%%%%%%%%%%%%%%%%%%%
%
  Obviously, one can reach a definite 
conclusion concerning the type of the spectrum
only if the value of $\meff_{\rm exp}$ 
is larger than $\meff_{\rm min}^{\rm QD}$, or is 
smaller than $\meff_{\rm max}^{\rm IH}$.
\vspace{-0.3cm}
\begin{itemize}
%%%%%%%%%%%%%%%%%%%%%%%%%%%%%
\item[i)] $\meff_{\rm exp} > \meff_{\rm min}^{\rm QD}$: 
this is equivalent to ruling out the IH spectrum and thus to the case 
i) analyzed in Subsection \ref{sec:outIH}, 
see eq.\ (\ref{eq:yIH}) and the discussion thereafter.  
\vspace{-0.3cm}
\item[ii)] $\meff_{\rm exp} < \meff_{\rm max}^{\rm IH}$:
using eqs.\ (\ref{eq:IHminmax}) and (\ref{eq:zeta}), 
we find that 
$\meff_{\rm exp} < \meff_{\rm max}^{\rm IH}$ if
%%%%%%%%%%%%%%%%%%%%%%%%%%%%%%%
\begin{equation} \label{eq:zetaIHQD}
1 < \zeta < \frac{(1 - s^2) \, \sqrt{(\dma\!\!)_{\rm max}}}
{(\meff_{\rm exp})_{\mbox{}_{\rm MIN}} + \Delta}~.   
\end{equation}
%%%%%%%%%%%%%%%%%%%%%%%%%%%%%%%%%
%
This inequality practically coincides with the second condition
in (\ref{eq:outIH4ab}).   
It is more restrictive
for smaller values of $(\dma\!\!)_{\rm max}$ 
and larger values of $\Delta$. Equation (\ref{eq:zetaIHQD}) 
can hold only for a rather limited 
range of parameters, since the sum of 
$(\meff_{\rm exp})_{\mbox{}_{\rm MIN}}$ 
and $\Delta$ has to be smaller than 
$\sqrt{(\dma\!\!)_{\rm max}} \simeq 0.07$ eV. 
\end{itemize}

\vspace{-0.2cm}
 Let us note that the various conditions 
discussed in this Section do not require 
any additional 
input from $^3$H $\beta$-decay experiments or 
from cosmological and astrophysical 
measurements.

%%%%%%%%%%%%%%%%%%%%%%%%%%%%%%%%%%%%%%%%%%%%%%%
\vspace{-0.5cm}
\section{\label{sec:parities}Distinguishing
Between Different Neutrino CP-Parity Configurations}
\vspace{-0.2cm}
%%%%%%%%%%%%%%%%%%%%%%%%%%%%%

\hskip 1.0truecm In this Section we will discuss 
whether a measurement of $\meff \neq 0$ 
might allow one to distinguish 
between some of the possible
neutrino CP-parity configurations
when the Majorana phases take CP-conserving values,
$\alpha_{21}, \alpha_{31} = 0, \pm \pi$.
We will denote these configurations by
$i^{-1}\left(\eta_{CP}(\nu_1)~\eta_{CP}(\nu_2)~\eta_{CP}(\nu_3) \right)$,
where $\eta_{CP}(\nu_j)$ is the CP-parity of the neutrino $\nu_j$,
$\eta_{CP}(\nu_j) = \pm i$. 
The possibility of determining the values
of the Majorana CP-violating phases 
in the general case of CP-non conservation
has been discussed in detail in ref.~\cite{PPR1}.

   Inspecting Tables $1 - 3$ leads 
to the conclusion that 
it might be relatively easy to distinguish 
between the \pmm{} and \mpm{} configurations 
in the case of the
IH spectrum (i.e., $\meff_{\rm min}^{\rm IH}$) and the 
\pmm{} and \mpm{} configurations for the QD spectrum (i.e., 
$\meff_{\rm min}^{\rm QD}$). 
The more interesting question is whether
it might be possible to distinguish
between the different CP-parity configurations
for a given type of neutrino mass spectrum.
We will study it briefly in what follows in
the cases of IH and QD spectra
\footnote{For an analysis of this possibility 
for the NH spectrum without taking into account 
the nuclear matrix element uncertainty, 
see \cite{PPW,WR0203}.}. 

%%%%%%%%%%%%%%%%%%%%%%%%%%%%%%%%%%%%%
\vspace{-0.4cm}
\subsection{\label{sec:paIH}Inverted Hierarchical Spectrum}
\vspace{-0.2cm}
%%%%%%%%%%%%%%%%%%%%%%%%%%%%%%%%%%%%%

\hskip 1.0truecm Due to the smallness of $m_1 \, |U_{e1}|^2$, 
one cannot distinguish 
the \ppp{} from the \mmp, as well as the \pmm{} from the \mpm,
configurations \cite{WR0203}. 
The first pair of CP-parity 
configurations corresponds to 
$\meff_{\rm max}^{\rm IH}$, while the second 
corresponds to $\meff_{\rm min}^{\rm IH}$.  
The CP-parity patterns
\ppp, \mmp{} and \pmm, \mpm{}
would be distinguishable if
the following condition holds:
%%%%%%%%%%%%%%%%%%%%%%%%%%%%%%%%%%%%%%%%%%%%%
\begin{equation}
\meff_{\rm max}^{\rm IH} > \zeta \, \meff_{\rm min}^{\rm IH}~. 
\end{equation}
%%%%%%%%%%%%%%%%%%%%%%%%%%%%%%%%%%%%%%%%%%%%
%
This can be translated into a condition on $\zeta$, which reads 
%%%%%%%%%%%%%%%%%%%%%%%%%%%%%%%%%%%%%%%
\begin{equation} \label{eq:zetaparIH}
\zeta < 
\frac{\sqrt{(\dma\!\!)_{\rm min}}}{\sqrt{(\dma\!\!)_{\rm max}}} \; 
\frac{1 + (\ts)_{\rm min}}{1 - (\ts)_{\rm min}}~.
\end{equation}
%%%%%%%%%%%%%%%%%%%%%%%%%%%%%%%%%%%%%%%
% 
The first ratio in the right-hand side of 
eq.\ (\ref{eq:zetaparIH}) is, 
for an assumed  error on \dma of 10 $\%$, approximately 
0.8, while for $(\ts)_{\rm min} = $ 0.2, 0.3, 0.4, 0.5, 0.6,
the second ratio reads 1.5, 1.9, 2.3, 3.0 4.0, respectively.
If $\zeta \ls 1.5$, 2.0, 3.0, 4.0,
values of $\ts \gs  0.3$, 0.4, 0.5, 0.7 
are required in order 
\newpage
\noindent to be possible to distinguish 
between the two cases under study. 

%%%%%%%%%%%%%%%%%%%%%%%%%%%%%%%%%%%%%%
\vspace{-0.4cm}
\subsection{\label{sec:paQD}Quasi-Degenerate Spectrum}
\vspace{-0.2cm}
%%%%%%%%%%%%%%%%%%%%%%%%%%%%%%%%%%%%%%

\hskip 1.0truecm In the case of QD spectrum,
the \pmm{} and \mpm, and the \ppp{} and \mmp, 
configurations are 
difficult to distinguish
due to the smallness of the mixing parameter 
$s^2$ limited by the
reactor antineutrino 
experiments \cite{CHOOZ,PaloV}: 
the corresponding differences in the 
predicted values of \meff{}
do not exceed $\sim 10 \%$. 
Therefore we shall analyze again 
the possibility to discriminate between
these two pairs. 
Taking into account that 
$\meff_{\tiny \pmm}^{\rm QD}  \gs \meff_{\tiny \mpm}^{\rm QD}$ and 
$\meff_{\tiny \ppp}^{\rm QD}  \gs \meff_{\tiny \mmp}^{\rm QD}$, 
the indicated two pairs of CP-parity 
configurations 
can be distinguished if 
the following inequality holds: 
%%%%%%%%%%%%%%%%%%%%%%%%%%%%%%%
\begin{equation}
\meff_{\tiny \mmp}^{\rm QD} \! > \zeta \, \meff_{\tiny \mpm}^{\rm QD}~.
\end{equation}
%%%%%%%%%%%%%%%%%%%%%%%%%%%%
%
The above inequality leads to the condition
%%%%%%%%%%%%%%%%%%%%%%%%%%%
\begin{equation}
\zeta <
\frac{(m_0)_{\rm min}}{(m_0)_{\rm max}} \; 
\frac{1 + (\ts)_{\rm max} \; (1 - s^2)}{1 - (\ts)_{\rm min} \; (1 - s^2)}~,
\label{eq:zetaparQD}
\end{equation}
%%%%%%%%%%%%%%%%%%%%%%%%%
%
which is very similar to eq.\ (\ref{eq:zetaparIH}). 
Assuming a KATRIN 
inspired error of 0.28 eV on $m_0$, 
the first fraction in the right-hand side of 
eq.\ (\ref{eq:zetaparQD}) 
is 0.1, 0.3, 0.6, 0.7 for 
$m_0 = $ (0.35, 0.5, 1.0, 1.5) eV, respectively. 
If $\sigma(3 \, m_0)= 0.10$ eV, as is 
expected from combined astrophysical 
and cosmological measurements, then 
for the same fraction one gets
0.83, 0.93, 0.94, 0.96 for 
$m_0 = $ (0.35, 0.5, 1.0, 1.5) eV, respectively. 
In what regards the second fraction, 
for $s^2 = 0$ the values from Section 
\ref{sec:paIH} are valid, while 
for $s^2 = 0.05$ they read 
1.7, 2.1, 2.6, 3.3 for 
$(\ts)_{\rm min} = $ 0.3, 0.4, 0.5, 0.6.
 
  Thus, if $m_0$ is measured in 
tritium $\beta$-decay 
experiments, relatively 
large $m_0 \gs 1.5$ eV and $\ts \gs 0.5$ 
are required in order to distinguish between
the \pmm, \mpm, and the \ppp, \mmp{} 
CP-parity configurations.
If astrophysical and cosmological 
measurements provide $m_0$, 
then  a value of $\zeta \ls $ 1.5, 2.0, 3.0, 4.0
would require $\ts \gs $ 0.2, 0.4, 0.5, 0.6.  

%%%%%%%%%%%%%%%%%%%%%%%%%%%%%
\vspace{-0.5cm} 
\section{Conclusions}
\vspace{-0.2cm}
%%%%%%%%%%%%%%%%%%%%%%%%%%%%

\hskip 1.0truecm 
  Assuming 3-neutrino mixing and 
massive Majorana neutrinos,
\betabeta-decay induced only by 
the (V-A) charged current weak interactions,
LMA MSW solution of the solar neutrino problem and
neutrino oscillation explanation 
of the atmospheric neutrino data,
we have studied the requirements on 
the ``solar'' mixing angle $\theta_\odot$, 
the nuclear matrix element uncertainty factor $\zeta$
and the experimental error on the
effective Majorana mass \meff, $\Delta$,
which allow one to distinguish between, 
and/or test, the normal hierarchical (NH), 
inverted hierarchical (IH) and 
quasi-degenerate (QD) neutrino 
mass spectra if $\meff \neq 0$ is measured, 
or a stringent upper bound on \meff{} is obtained. 
The possibility to discriminate
between the three types of spectra
depends on the allowed 
ranges of values of \meff{} for the 
three spectra: it is determined by 
the maximal values of \meff{} in the cases of NH and IH spectra,
$\meff_{\rm max}^{\rm NH,IH}$,
and by the minimal values of \meff{} 
for the IH and QD spectra,
$\meff_{\rm min}^{\rm IH,QD}$.
These  are reported in Tables $1-3$. In deriving 
them we have used the 
values of the solar and atmospheric 
neutrino oscillation parameters,
$\theta_\odot$, \deltasol, \dma$\!\!$ and 
$\sin^2\theta$,   
favored by the existing 
data \cite{Cl98,SKsol,SNO2,SKatm00,KamLAND,CHOOZ,PaloV}
(Tables 1 and 2) and 
assumed prospected precisions of 
future measurements (Table 3).

   For the currently favored 
values of the neutrino oscillation 
parameters and $\sin^2\theta = 0$,  
the upper bound on \ts{} permitting to 
distinguish the NH from the IH spectrum  
is satisfied even for 
$\zeta = 3$. 
If $\sin^2\theta$ lies close to its present
99.73\% C.L. upper limit of 0.05 
\cite{fogli} (see also
\cite{CHOOZ,PaloV}), 
the upper bound of interest 
decreases by up to 50\%
and values of $\zeta$ 
slightly lower than 3
might be required (Fig. 1). 
The possibility to discriminate 
between the NH and the QD spectra   
depends weakly on $sin^2\theta$ and on
the neutrino mass $m_{1,2,3} \cong m_0$, and 
the respective conditions are
satisfied even for values of 
$\zeta$ exceeding 3 (Fig.\ 1). 
Without any additional input 
from $^3$H beta-decay experiments and/or
cosmological and astrophysical 
measurements, and given the 
values of $\ts$ and $\dma$ 
favored by the data, the IH and QD spectra 
can be distinguished only 
if $\zeta \ls 1.5$ (Fig.\ 2). 

  Let us emphasize that the conditions
which would allow one to establish
the presence of CP violation 
due to the Majorana CP-violating 
phases using a measurement of
$\meff \neq 0$ lead to 
a lower bound on $\ts$ 
and, in general, require $\zeta < 2$. 
In contrast, the conditions 
permitting to distinguish between
the three types of neutrino mass spectrum
imply an {\it upper limit} on $\ts$ and
in most of the cases can be satisfied 
even for $\zeta \simeq 3$.

   We have studied also 
the conditions on $\zeta$ and $\Delta$
which would permit to rule out, or establish,
the NH, IH and the QD mass spectra. 
Typically, the next generation of 
\betabeta-decay experiments will be 
able to rule out the QD mass spectrum 
if $\zeta \ls 3$, and establish it 
if, e.g., the measured  
$(\meff_{\rm exp})_{\mbox{}_{\rm MIN}}\sim 0.2~(0.1)$ eV
(see eq. (\ref{eq:zeta})) 
and the experimental error is 
$\Delta \sim 0.15~(0.05)$ eV\@. 
The NH spectrum can be excluded 
provided the measured value of \meff{} 
is, e.g,  $\sim 10~(7)$ times 
larger than $\meff_{\rm max}^{\rm NH}$ and 
the experimental error is $\Delta \ltap 0.12~(0.06)$ eV\@. 
The IH spectrum can be ruled out 
for $\Delta \ltap 0.07~(0.10)$ eV
provided 
$(\meff_{\rm exp})_{\mbox{}_{\rm MIN}}$
is at least by a factor
of $\sim 2.0~(2.5)$ larger  than
$\meff_{\rm max}^{\rm IH}$. 
Establishing the IH mass spectrum
is quite demanding and
requires a measurement 
of \meff{} with an error 
$\Delta \ltap 0.02 - 0.04$ eV\@. 
  
   Finally, we have studied 
the possibility to distinguish between certain 
neutrino CP-parity configurations in the case
of CP-conservation. 
Due to the smallness of $\sin^2\theta$,
there are two pairs of CP-parities
in the cases of QD and IH spectra,
the two different 
CP-parity patterns 
within each pair
being indistinguishable.
Given the best-fit values of $\ts$, 
one can discriminate between 
these two pairs for the IH 
mass spectrum if $\zeta \ls 2$. 
For the QD mass spectrum and 
if $m_0$ is measured in 
tritium $\beta$-decay 
experiments, relatively 
large $m_0 \gs 1.5$ eV and $\ts \gs 0.5$ 
are necessary.  
If astrophysical and cosmological 
measurements provide $m_0$, 
values of $\zeta \ls 2$ are required.  

\vspace{0.2cm}
{\bf Acknowledgments.}
Part of the present work
was done during the Workshop on $\betabeta-$Decay 
at the Institute of Nuclear Theory (ITP) at 
the University of Washington, Seattle.
S.P. and S.T.P. would like to thank 
the organizers of the Workshop 
and the members of ITP for kind hospitality and 
L. Wolfenstein, W. Haxton, P. Vogel, 
R. Mohapatra and F. Avignone
for useful discussions.
This work was supported in part by 
the EC network HPRN-CT-2000-00152
(S.T.P. and W.R.), the Italian 
MIUR under the program ``Fenomenologia delle 
Interazioni Fondamentali'' (S.T.P.)
and by the U. S. Department of Energy (S.P.).

\newpage
\begin{table}[ht]
\begin{center}
\begin{tabular}{|c|c|c|c|c|c|c|} 
\hline
\rule{0pt}{0.5cm} Reference & $(\ts)_{\rm BF}$ & $(\dms\!\!)_{\rm BF}$  
& $\meff_{\rm max}^{\rm NH}$ & $\meff_{\rm min}^{\rm IH}$ & 
$\meff_{\rm min}^{\rm QD} $  
\\ \hline \hline
% Fogli {\it et al.} 
\cite{fogli} & 0.46 & 7.3  
&  5.9 & 18.4 & 59.9 \\ \hline
% Valle {\it et al.} 
\cite{valle} & 0.46 & 6.9  
&  5.8 & 18.4 & 59.9 \\ \hline
% Bandyopadhyay {\it et al.} 
\cite{band} & 0.42 & 7.2 
&  5.7 & 20.3 & 67.2 \\ \hline
% Bahcall {\it et al.} 
\cite{bahcall} & 0.43 & 7.0 
&  5.7 & 19.8 & 65.3 \\ \hline
\end{tabular}
\caption{\label{tab:BF} The best-fit values of \ts{} and \dms 
(in units of $10^{-5} \; \rm eV^2$) 
in the LMA solution region,
as reported by different authors. Given are 
also the calculated maximal values of  
\meff{} (in units of $10^{-3}$ eV) for the NH spectrum 
and the minimal values of 
\meff{} (in units of $10^{-3}$ eV) 
for the IH and QD spectra.
The results for \meff{} in the cases of 
NH and IH spectra are obtained for 
$m_1 = 10^{-3}$ eV and the 
best-fit value of \dma$\!\!$,
$\dma = 2.7 \times 10^{-3} \eV^2$ 
\cite{fogliold},
while those for 
the QD spectrum are derived for 
$m_0 = 0.2$ eV. 
In all cases $\sin^2\theta = 0.05$ 
has been used. The chosen value of
\dma corresponds to
$\meff_{\rm max}^{\rm IH} = 52.0 \times 10^{-3} \; \rm eV$.
}
\end{center}
\end{table}

\begin{table}[ht]
\begin{center}
\begin{tabular}{|c|c|c|c|c|c|c|} 
\hline
\rule{0pt}{0.5cm}Reference % Group 
& $\ts$ & \dms  
& $\meff_{\rm max}^{\rm NH}$ & $\meff_{\rm min}^{\rm IH}$ & 
$\meff_{\rm min}^{\rm QD} $  
\\ \hline \hline
% Fogli {\it et al.} 
\cite{fogli} & 0.32 $-$ 0.72 & 5.6 $-$ 17   
&  8.6 & 7.6 & 20.6 \\ \hline
% Valle {\it et al.} 
\cite{valle} & 0.31 $-$ 0.68 & 5.7 $-$ 15    
&  8.1 &  8.9 & 25.8 \\ \hline
% Bandyopadhyay {\it et al.} 
\cite{band} & 0.31 $-$ 0.56 & 6.0 $-$ 8.7   
&  6.6 & 13.0 & 43.2 \\ \hline
% Bahcall {\it et al.} 
\cite{bahcall} & 0.31 $-$ 0.66 & 5.9 $-$ 8.9   
&  7.0 &  9.5 & 28.6 \\ \hline
\end{tabular}
\caption{\label{tab:90} 
The ranges of allowed values of \ts{} and \dms 
(in units of $10^{-5} \; \rm eV^2$) 
in the LMA solution region, 
obtained at 90$\%$ C.L.\ by different authors.
Given are also the corresponding maximal values of  
\meff{} (in units of $10^{-3}$ eV)
for the NH spectrum, and the minimal values of 
\meff{} (in units of $10^{-3}$ eV) for the IH and QD spectra.
The results for the NH and IH spectra 
are obtained for $m_1 = 10^{-3}$ eV, while
those for the QD spectrum correspond to 
$m_0 = 0.2$ eV\@. % The value of 
~\dma was assumed
to lie in the interval \cite{fogliold}
$(2.3 - 3.1) \times 10^{-3} \eV^2$. This implies 
$\meff_{\rm max}^{\rm IH} = 55.7 \times 10^{-3} \; \rm eV$.
As in Table 1, the value $\sin^2\theta = 0.05$ was used.}
\end{center}
\end{table}

\begin{table}[ht]
\begin{center}
\begin{tabular}{|c|c|c|c|c|} 
\hline
\rule{0pt}{0.5cm} Reference % Group   
& $\meff_{\rm max}^{\rm NH}$ & $\meff_{\rm min}^{\rm IH}$ & 
$\meff_{\rm min}^{\rm QD} $  
\\ \hline \hline
% Fogli {\it et al.} 
\cite{fogli} 
& 6.1~(6.7) & 16.5~(12.9) &  55.9~(48.2) \\ \hline
% Valle {\it et al.} 
\cite{valle}
& 6.1~(6.6)  & 16.5~(12.9) & 55.9~(48.2) \\ \hline
% Bandyopadhyay {\it et al.} 
\cite{band}
& 6.0~(6.5)  & 18.3~(14.6) & 63.3~(55.9) \\ \hline
% Bahcall {\it et al.} 
\cite{bahcall}
& 6.0~(6.5) & 17.9~(14.1) & 61.4~(54.0)  \\ \hline
\end{tabular}
\caption{\label{tab:fut1} The values of 
$\meff_{\rm max}^{\rm NH}$, $\meff_{\rm min}^{\rm IH}$ and 
$\meff_{\rm min}^{\rm QD}$ 
(in units of $10^{-3}$ eV), calculated using   
the best-fit values of solar and atmospheric neutrino
oscillation parameters from Table \ref{tab:BF} 
and including 1 s.d.\ (3 s.d) 
uncertainties of 5 $\%$ (15\%)
on \ts{} and $\dms\!\!$, and
of 10 $\%$ (30\%) on \dma$\!\!$.
In this case one has: 
$\meff_{\rm max}^{\rm IH} = 54.5~(59.2) \times 10^{-3} \; \rm eV$.}
\end{center}
\end{table}

\newpage
\clearpage

\clearpage

\end{document}